\documentclass[manuscript,screen]{acmart}
\usepackage{amsfonts}
\usepackage{algpseudocode}
\usepackage{algorithm}
\usepackage{array}
\usepackage[caption=false,font=normalsize,labelfont=sf,textfont=sf]{subfig}
\usepackage{textcomp}
\usepackage{stfloats}
\usepackage{url}
\usepackage{verbatim}
\usepackage[justification=justified]{caption}
\usepackage{float}
\usepackage{booktabs}
\usepackage{xcolor}
\usepackage{amsmath}

\AtBeginDocument{%
  \providecommand\BibTeX{{%
    \normalfont B\kern-0.5em{\scshape i\kern-0.25em b}\kern-0.8em\TeX}}}

\setcopyright{acmlicensed}
\copyrightyear{2024}
\acmYear{2024}
\acmDOI{XXXXXXX.XXXXXXX}

\begin{document}

\title{SIDQL: An Efficient Keyframe Extraction and Motion Reconstruction Framework in Motion Capture}

\author{Xuling ZHANG}
\authornote{Both authors contributed equally to this research.}
\email{xzhang659@connect.hkust-gz.edu.cn}
\orcid{1234-5678-9012} %not change
\author{Ziru ZHANG}
\authornotemark[1]
\email{zzhang758@connect.hkust-gz.edu.cn}
\affiliation{%
  \institution{The Hong Kong University of Science and Technology (Guangzhou)}
  %\streetaddress{P.O. Box 1212} %not change
  \city{Guangzhou}
  \country{China}
  \postcode{511453}
}

\author{Yuyang WANG}
\affiliation{%
 \institution{The Hong Kong University of Science and Technology (Guangzhou)}
 %\streetaddress{Rono-Hills}
 \city{Guangzhou}
 \country{China}}
 \email{yuyangwang@hkust-gz.edu.cn}

\author{Lik-Hang LEE}
\affiliation{%
 \institution{The Hong Kong Polytechnic University}
 %\streetaddress{Rono-Hills}
 \city{Hong Kong}
 \country{China}}
 \email{lik-hang.lee@polyu.edu.hk}

\author{Pan HUI}
\affiliation{%
 \institution{The Hong Kong University of Science and Technology (Guangzhou)}
 %\streetaddress{Rono-Hills}
 \city{Guangzhou}
 \country{China}}
 \email{panhui@ust.hk}

\renewcommand{\shortauthors}{Zhang and Zhang, et al.}

\begin{abstract}
  Metaverse, which integrates the virtual and physical worlds, has emerged as an innovative paradigm for changing people's lifestyles. Motion capture has become a reliable approach to achieve seamless synchronization of the movements between avatars and human beings, which plays an important role in diverse Metaverse applications. However, due to the continuous growth of data, current communication systems face a significant challenge of meeting the demand of ultra-low latency during application. In addition, current methods also have shortcomings when selecting keyframes, e.g., relying on recognizing motion types and artificially selected keyframes. Therefore, the utilization of keyframe extraction and motion reconstruction techniques could be considered a feasible and promising solution. In this work, a new motion reconstruction algorithm is designed in a spherical coordinate system involving location and velocity information. Then, we formalize the keyframe extraction problem into an optimization problem to reduce the reconstruction error. Using Deep Q-Learning (DQL), the Spherical Interpolation based Deep Q-Learning (SIDQL) framework is proposed to generate proper keyframes for reconstructing the motion sequences. We use the CMU database to train and evaluate the framework. Our scheme can significantly reduce the data volume and transmission latency compared to various baselines while maintaining a reconstruction error of less than 0.09 when extracting five keyframes.

\end{abstract}

%%在网站自动生成，关键词/研究领域

\begin{CCSXML}
<ccs2012>
   <concept>
       <concept_id>10010147.10010257.10010258.10010261</concept_id>
       <concept_desc>Computing methodologies~Reinforcement learning</concept_desc>
       <concept_significance>500</concept_significance>
       </concept>
   <concept>
       <concept_id>10010147.10010371.10010352.10010380</concept_id>
       <concept_desc>Computing methodologies~Motion processing</concept_desc>
       <concept_significance>300</concept_significance>
       </concept>
   <concept>
       <concept_id>10010147.10010178.10010224.10010226.10010238</concept_id>
       <concept_desc>Computing methodologies~Motion capture</concept_desc>
       <concept_significance>500</concept_significance>
       </concept>
   <concept>
       <concept_id>10010147.10010178.10010224.10010226.10010239</concept_id>
       <concept_desc>Computing methodologies~3D imaging</concept_desc>
       <concept_significance>300</concept_significance>
       </concept>
 </ccs2012>
\end{CCSXML}

\ccsdesc[500]{Computing methodologies~Reinforcement learning}
\ccsdesc[300]{Computing methodologies~Motion processing}
\ccsdesc[500]{Computing methodologies~Motion capture}
\ccsdesc[300]{Computing methodologies~3D imaging}

\keywords{Keyframe Extraction, Motion Reconstruction, Deep Q-Learning, Spherical Interpolation.}

%I delete them

\maketitle

\section{Introduction}
%The Metaverse is a virtual world that exists in parallel to the physical world, created by various technologies such as artificial intelligence (AI), virtual reality (VR), augmented reality (AR) and human-machine interaction (HMI). It has high application value and is widely used in games, education, medical, military and other fields \cite{metaverse_01}. In addition, the Metaverse is changing the way of communication among human beings. VR and AR devices enable users to interact in an immersive environment, providing a more engaging experience than a flat environment would allow. It can improve communication efficiency among users and between users and machines~\cite{metaverse_02}. Avatar is an important concept in the Metaverse since every user needs an avatar in the virtual world for interaction. It is a 3D digital figure with human-like qualities created through convergence technology, which is the main carrier of virtual space-time perception for natural persons in the Metaverse. Users in the future have unique avatars and rely on the avatar to enter the virtual world to synchronize real-world motions for exploration \cite{avatar_01, avatar_02}.

The Metaverse, a virtual realm that coexists alongside the tangible world, is forged by a synergy of technologies, including artificial intelligence (AI), virtual reality (VR), augmented reality (AR), and human-machine interaction (HMI). Esteemed for its high application potential, it finds extensive use across various domains such as gaming, education, healthcare, and defense \citep{metaverse_01}. Moreover, the Metaverse is revolutionizing human communication modalities. Through VR and AR devices, users engage in a deeply immersive environment that surpasses the engagement offered by traditional, two-dimensional settings. This advancement fosters enhanced communicative efficiency both amongst users and between users and machines \cite{metaverse_02}. Within the Metaverse, avatars stand as a pivotal concept, with each user adopting a digital persona for virtual interactions. These avatars, three-dimensional digital entities endowed with anthropomorphic traits via convergent technologies, serve as the principal conduit for an individual's perception of virtual temporality within the Metaverse. Looking ahead, users will possess distinctive avatars, leveraging these virtual proxies to mirror real-life movements within the exploratory confines of the virtual domain \cite{avatar_01, avatar_02}.

Motion capture technology is the predominant method for synchronizing avatars' movements in the Metaverse. By capturing the movements of actors or performers, motion capture enables the creation of avatars that mimic their actions, transporting users into a world where they can seamlessly interact with these digital representations. With motion capture techniques, avatars enable users to embark on immersive journeys, participate in social interactions, and even engage in exhilarating 3D games, which plays a pivotal role in the development of the Metaverse \cite{application_02}. In addition, the combination of motion capture, 3D animations, and 3D games presents an unprecedented opportunity to bridge the gap between the real and digital worlds, pushing the boundaries of imagination and unleashing a new era of interactive and immersive experiences \cite{application_01}.

The naive approach of synchronizing an avatar's movements in the Metaverse requires transmitting a large amount of motion data in a short time, unavoidably causing perceptible delays in avatar movements. So, keyframe extraction and motion reconstruction techniques can be utilized to reduce the latency. Keyframe extraction techniques, commonly used in videos, are valuable as they summarize video content by identifying a concise set of representative frames. It can reduce the amount of motion data that needs to be transmitted to the users. There has been a lot of research on video-based keyframe extraction these years, which provides valuable insights and techniques that enhance video storage, transmission and summarization \cite{video_ke_01, video_ke_02}. Due to the different data formats, keyframe extraction in motion capture sequences takes a distinct approach compared to video content. Keyframe extraction in motion capture involves selecting a set of representative frames, which correspond to the positions of human skeletal points, to represent the overall motion sequence effectively. Keyframes represent the significant poses or positions captured during the motion capture process \cite{subspace_learning}. Therefore, we can transmit the keyframes of motion data in situations of network fluctuations, effectively meeting the requirement for low latency in the Metaverse.

However, we cannot directly show the extracted keyframes on the user's device after keyframe extraction and transmission, as the transitions between these keyframes may appear sudden or stiff. To enhance smooth video playback for users, advanced video technologies utilize frame reconstruction techniques after the frame reduction \cite{video_reconstruct_01, video_reconstruct_02}. Similar research exists in the field of motion capture. Motion reconstruction in motion capture is to generate natural in-between motions in terms of a given start-end keyframe pair \cite{reconstruction_snerp}. It can be utilized to fill this gap by creating coherent and realistic in-between motions that smoothly connect the keyframes, enhancing the overall quality of the captured motion.

Thus, novel motion capture techniques are crucial in achieving realistic and synchronized avatars' movements in the Metaverse. As virtual environments become increasingly immersive, it is essential to accurately capture and map human motion in the real world to the virtual world. Since human motion data is usually captured at high frequency, the huge amount of motion data increases transmission consumption, resulting in a delay in avatar movements. However, avatars' movements in the Metaverse have to meet low latency requirements. Therefore, keyframe extraction and motion reconstruction in motion capture become vital challenges to meet the low latency requirement. With these two methods, we are able to transmit only the keyframes of motion data despite the fluctuating network, and then reconstruct the frames on the user side, achieving movement synchronization between avatars and humans.

According to the existing work, we find that current AI-based methods always rely on labeled data for training. However, the accuracy of the labeled keyframes is uncertain as the labeled keyframes are based only on human intuition rather than selecting the most suitable keyframes. Since DQL can solve this problem, opportunities exist for designing a new keyframe extraction algorithm on top of DQL without using labeled dataset for training.

In parallel, some motion reconstruction methods ignore the constancy of bone length, resulting in incoherent reconstructed movements. Also, some approaches need to identify the motion category before the motion can be reconstructed, which makes the motion reconstruction limited by the motion category and takes a long time to reconstruct. Moreover, velocity information of human bones is often disregarded in current motion reconstruction methods, causing the underutilization of information in non-keyframes. Inspired by the above-mentioned articles and algorithms, the article aims to investigate a new motion reconstruction method that fully utilizes the velocity information to improve the quality of reconstructed motion. We propose a new motion reconstruction algorithm considering the information hidden in the non-keyframes. The proposed reconstruction algorithm, primarily a mathematical model and DQL, serves to decide the keyframe extraction optimally. Finally, the model is evaluated via experiments using motion data in the mixed category. The major contributions of this paper can be summarized as follows:

\begin{itemize}
    \item We first develop a new motion reconstruction algorithm in a spherical coordinate system. The location and velocity data of each point are converted into a spherical coordinate system to keep the length of the bones constant. Then, a spherical interpolation approach is conducted to reconstruct the middle frames. The motion of the root point is also reconstructed by utilizing polynomial interpolation methods.
  %We formalize the reconstruction problem and designed a new motion reconstruction algorithm. We use polynomial interpolation method to reconstruct the motion of the root point, which allows us to reconstruct the motion trajectory without identifying the motion category. To keep the bone length constant, the motion of other skeletal points is reconstructed based on spherical interpolation. By using the new motion reconstruction algorithm, we can reconstruct the non-keyframes with better accuracy.
    \item To minimize the mean reconstruction error, we formalize the keyframe extraction problem into an optimization problem according to the motion reconstruction method. A SIDQL framework for keyframe extraction and motion reconstruction is then proposed by using a special reward function developed based on mean error. The SIDQL framework can be trained by mixed-category motion sequences without labeled keyframes.

    \item We then conduct comprehensive experiments to study the impact of different hyperparameters. The reconstruction performance is measured by comparison analysis with various baselines under various numbers of keyframes. The experiments give evidence that our scheme can significantly reduce the data volume with promising reconstruction accuracy.
      %We conduct several experiments by using the motion data in mixed category. Then we evaluate the proposed approach, which achieves superior performance compared with different methods.
\end{itemize}

\section{RELATED WORK}

Keyframe extraction and motion reconstruction are foundation techniques for various fields, such as animation, gaming, and medical research. To reduce data redundancy and improve transmission efficiency, researchers proposed various keyframe extraction models in video and motion capture scenarios. In addition, motion reconstruction approaches are widely applied to enhance the visual quality in video sequences or motion capture data. This section briefly summarizes some previous studies about different keyframe extraction models and motion reconstruction approaches.

\subsection{Keyframe Extraction}

Keyframe extraction plays a significant role in both video and motion capture domains. The extracted keyframes can reduce data redundancy, enhance transmission efficiency and provide meaningful representations for various applications, including video summarization, motion analysis and animation synthesis.

Keyframe extraction in videos refers to the process of selecting a subset of frames that represent the essential content or significant moments within the video sequence. Some researchers use clustering methods to extract keyframes in videos. For example, Guru et al. \cite{video_cluster_01}, and Jadon et al. \cite{video_cluster_02} utilized Gaussian Mixture Model and K-means clustering method to cluster all frames based on features, where the cluster centers represented the keyframes. In addition to clustering algorithms, deep learning is also commonly used for video keyframe extraction. For example, Ly et al. \cite{video_dl_01} first utilized the Convolutional Neural Network (CNN) and Iterative Quantization (ITQ) method to calculate the hamming distance of each frame. If the hamming distance differed from the previous frame, it was considered as a keyframe. In \cite{video_dl_02}, CNN and Recurrent Neural Network (RNN) were used to identify the information regions of each video frame. The frames with higher structural similarity scores were considered as keyframes. Furthermore, keyframe extraction in videos has widespread applications in the field of medical screening. For instance, Huang et al. \cite{video_medical_01} introduced a reinforcement learning-based framework for extracting keyframes in breast ultrasound videos automatically, while Pu et al. \cite{video_medical_02} utilized CNN and RNN to select keyframes in fetal ultrasound videos.

With the growing demand for realistic and high-quality animations in various industries, keyframe extraction in motion capture attracts increasing attention, which is the process of identifying and extracting significant frames from a motion sequence. The curve simplification-based approach is an important kind of keyframe extraction method \cite{mocap_curve_01, mocap_curve_02}. In this method, a single frame from a motion sequence can be viewed as a point within a high-dimensional space, and these points can collectively form a trajectory curve. The keyframes are then selected as the junctions between the simplified curve segments. Bulut et al. \cite{mocap_curve_01} proposed a curve saliency metric that extracts keyframes from motion curves, where points with saliency values higher than the average saliency value were identified as keyframes. Zhang et al. \cite{mocap_curve_02} employed the Principal Component Analysis (PCA) method to derive feature curves from a motion sequence, and subsequently identified the keyframes by extracting local optimum points along the curves. Another alternative method for keyframe extraction is the clustering-based approach, where similar frames are grouped into clusters. From each cluster, a representative frame is selected as a keyframe for the motion sequence. For instance, Zhang et al. \cite{mocap_cluster_01} introduced the ISODATA method for clustering all frames in a motion sequence, with the frames closest to the center of each cluster being identified as keyframes. Sun et al. \cite{mocap_cluster_02} proposed an affine propagation clustering algorithm to cluster all the frames, where the center of each cluster was chosen as the keyframe. Furthermore, some researchers apply matrix factorization-based methods to extract the keyframes. They utilize a motion matrix to represent a motion sequence, which is subsequently decomposed into a weight matrix and a keyframe matrix. Huang et al. \cite{mocap_matrix_01} designed the Key Probe method to factorize the motion matrix into two smaller matrixes, then used the least-squares optimization method to solve the keyframe extraction problem.

In recent years, AI-based keyframe extraction methods have also attracted interest from researchers. Genetic Algorithm (GA) is an important heuristic algorithm for decision-making problems. In \cite{mocap_ga_01}, Liu et al. combined GA with a probabilistic simplex method to propose a keyframe extraction algorithm called Simplex Hybrid Genetic Algorithm (SMHGA), which enhanced the data processing speed compared to GA. The keyframes were obtained through an iterative application of the algorithm, where initial populations were generated randomly and intelligently. Zhang et al. \cite{mocap_ga_02} presented a Multiple Population Genetic Algorithm (MPGA) to extract the keyframes of motion capture data by minimizing the reconstruction error. While the SMHGA utilized a single group for evolution, the MPGA employed multiple populations, resulting in a larger search space for the optimization process. Except for GA, Particle Swarm Optimization (PSO), as another prominent heuristic algorithm, can also be used for keyframe extraction. Chang et al. \cite{mocap_pso} designed a keyframe extraction method based on PSO. The iteration of PSO stopped when the variance was smaller than the set value, and then the output was considered as keyframes. However, the heuristic algorithms always take a long time to train. Some sparse represent-based methods \cite{sr_01, sr_02} can be used to solve the problem. Xia et al. \cite{mocap_jointkernelsr} proposed a joint kernel sparse representation model to extract the keyframes, which could simultaneously model the sparseness and the Riemannian manifold structure of the human skeleton. Also, some researchers use deep learning methods to solve the keyframe extraction problem. For example, Prabakaran et al. \cite{mocap_ocnn} first used PCA to reduce the data size, then proposed a new algorithm to extract the keyframes of motion data based on Optimized Convolution Neural Network (OCNN) and Intensity Feature Selection (IFS). A graph-based deep reinforcement learning method for keyframe extraction was proposed by Mo et al. \cite{mocap_dql}, in which each frame is represented as a graph.

From the above work, it is envisioned that researchers have done a lot of work on keyframe extraction. As human motion data is captured frame-by-frame, we are much inspired by some video-based approaches. However, these approaches are limited to extracting keyframes from images, which cannot be directly applied to select the keyframes in sequential motion data. Since the bone length is fixed, researchers use mathematical methods to calculate the joints, but these methods are computationally intensive. With the development of AI technology, some researchers use AI-based methods to reduce the computational effort of models. Although these methods are fast and accurate, some of them need to identify the type of motion before training, which is not widely applied. In addition, some AI-based methods convert each frame of motion data to a graph, which also requires high computational cost. Furthermore, existing AI-based methods rely heavily on labeled data for training, but the selected keyframes are labeled according to human intuition, which contains much subjective noise and uncertainty, so the accuracy of the labels is doubtful. As there are no fixed categories of avatar motions in the Metaverse and it is difficult to label keyframes artificially, it is a very challenging task to extract keyframes in the Metaverse.

\subsection{Motion Reconstruction}

In addition to keyframe extraction algorithms, there is still plenty of work that mainly focuses on the reconstruction of videos and motion capture. Reconstruction techniques aim to enhance the visual quality, resolution, and fidelity of individual frames in a video sequence or motion capture data, which can be used in many different areas.

Video reconstruction refers to the process of generating a new video sequence from a given input, typically with the goal of improving or enhancing the original video. Deep learning methods have recently demonstrated the potential to solve the video reconstruction problem. Weng et al. \cite{reconstruction_cnn_01} and Rebecq et al. \cite{reconstruction_cnn_02, reconstruction_cnn_03} proposed CNN-based models for video reconstruction, leveraging the inherent capability of CNNs to capture and learn intricate spatial and temporal patterns in video data. Wang et al. \cite{reconstruction_gan_01} reconstructed the frames from grayscale videos with Generative Adversarial Networks (GANs). In addition, video reconstruction has a wide range of applications, such as medical imaging \cite{reconstruction_medical} and multimedia \cite{reconstruction_multimedia}.

With the development of motion capture technologies, motion reconstruction has been a significant area of research, aiming to generate natural intermediate motions based on a given start-end keyframe pair. Various techniques and methodologies have been explored in this field. In the early stage, most of the motion reconstruction methods were designed with linear interpolation~\cite{mocap_curve_02, mocap_cluster_01, mocap_matrix_01, mocap_ga_02}, providing a simple and straightforward way for motion reconstruction. However, linear interpolation may result in incoherent motions, especially when significant variations exist between consecutive frames. Polynomial interpolation, on the other hand, offers more flexibility by fitting curves or polynomials to the motion sequences \cite{reconstruction_poly_01, reconstruction_poly_02}, enabling smoother and more accurate reconstruction of motion trajectories compared with linear interpolation. However, the bone length should be a constant value. Otherwise, the movements will appear incoherent and stiff. Both linear interpolation and polynomial interpolation change the length of the bones, resulting in lower accuracy of motion reconstruction. To solve this problem, the Spherical Linear Interpolation (Slerp) method has been explored to reconstruct the motion \cite{mocap_cluster_02, mocap_pso, reconstruction_slerp_01}, ensuring that the bone length remains constant during the interpolation. The Slerp method interpolates between two quaternions along the shortest path on the unit sphere, which maintains a constant angular velocity during the interpolation process. Additionally, motion reconstruction based on deep learning has gained significant attention in recent years. Xia et al. \cite{reconstruction_snerp} proposed a Sphere Nonlinear Interpolation (Snerp) model to reconstruct the motion based on paired dictionary learning and sphere interpolation. In contrast to Slerp, Snerp does not maintain a constant angular velocity during interpolation, resulting in motion that closely resembles human skeletal movement and thereby improving the accuracy of motion reconstruction. Kim et al. \cite{reconstruction_dnn} utilized a DNN with an attention mechanism to reconstruct the motion, which could effectively observe the long-term information with the attention layers.

The work mentioned above has demonstrated that researchers have made significant advancements in motion reconstruction in recent years. Few works have been done with linear interpolation and polynomial interpolation, but the reconstructed motion appears to be incoherent and stiff due to the non-fixed length of the bones. While the Slerp method effectively addresses this issue, its drawback lies in maintaining a constant angular velocity during interpolation, leading to motion that does not closely resemble human skeletal movement. Also, deep learning-based methods always rely on labeled data for training, which cannot be applied to motion synchronization scenarios. Nevertheless, this kind of method always needs to identify the type of motion before training, which cannot be widely applied. However, some motion capture devices can directly support the calculation of velocity information nowadays, further increasing the amount of motion information. As most of the existing motion reconstruction methods do not involve velocity, the information in non-keyframes is not fully exploited. Here, we propose a SIDQL framework for keyframe extraction and motion reconstruction. The polynomial interpolation method is used to reconstruct the motion of the root point, while a new spherical interpolation-based method is utilized to generate the motion of other skeletal points. In addition, DQL is used to extract the keyframes without using labeled data.

\section{PROBLEM FORMULATION}

To solve the challenges of keyframe extraction and motion reconstruction in motion capture, a new math model of the whole procedure is developed. We first convert the sequences into spherical coordinates and propose a new motion reconstruction method utilizing the location and velocity information. The keyframe extraction problem is then formalized into an optimization problem, which aims to minimize the reconstruction error. For convenience, the major notations used in this paper are listed in Table \ref{notation}.

\begin{table}[htbp]   
\renewcommand{\tablename}{TABLE}
\caption{Major Notations}   
\begin{center}
\resizebox{0.7\textwidth}{!}{%
\begin{tabular}{ll}    
\toprule    \bf{Notation} & \bf{Description}\\    
\midrule  

N & The number of frames in a sequence.\\
M & The number of points in a frame.\\
$F_n$ & The $n_{th}$ frame in sequence S.\\ 
$r_m$ & The length of the $m^{th}$ bone.\\
$\theta_m$ & The angle value between the vector of the $m^{th}$ bone and the z-axis.\\
$\phi_m$ & The angle value between the vector of the $m^{th}$ bone and the y-axis.\\
$(\theta_{n,m},\phi_{n,m})$ & The $m^{th}$ point in the $n_{th}$ frame. \\  
$(\dot{\theta}_{n,m},\dot{\phi}_{n,m})$ & The velocity of $m^{th}$ point in the $n_{th}$ frame. \\  
$(\Tilde{\theta}_{n,m},\Tilde{\phi}_{n,m})$ & The $m^{th}$ point in the $n_{th}$ reconstructed frame.\\
$\Delta$ & The time interval between two adjacent frames.\\
K & The keyframe extraction decision of sequence S.\\
$k_w$ & The index of the $w^{th}$ keyframe.\\
E & The mean error after reconstruction.\\

\bottomrule   
\end{tabular}  }
\end{center}
\label{notation}
\end{table}

\subsection{Motion Reconstruction}

\begin{figure}[htbp]
	\centerline{
	\includegraphics[scale=0.072,trim=0in 01in 0in 0.4in]{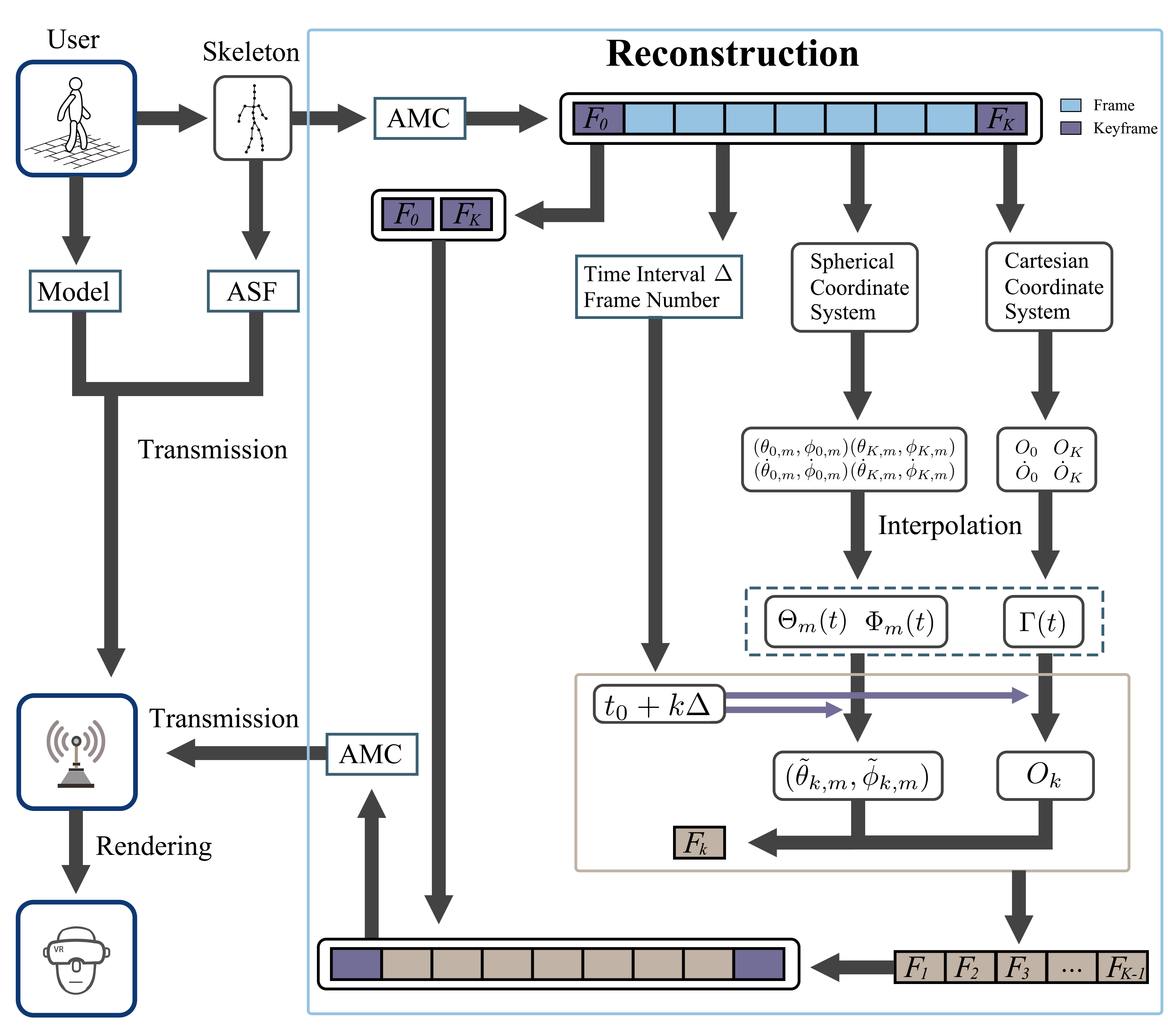}}
	\caption{The procedure of reconstruction.}
	\label{reconstruction}
\end{figure}

Motion reconstruction of motion capture is a problem that uses two frames in a sequence to reconstruct the middle frames via different approaches. By using reconstruction techniques, we increase the frame rate of the motion, improve the consistency of movement and reduce the size of transmitted data. Since the bone length is fixed, the traditional interpolation algorithm in the Cartesian coordinate system fails to ensure a constant length, which will greatly influence the watching experience. In addition, linear interpolation approaches will inevitably influence the smoothness between different sections. As a consequence, we proposed a new motion reconstruction algorithm with polynomial interpolation in the spherical coordinate system. The framework of motion reconstruction is shown in Fig \ref{reconstruction}. ASF files contain information about the skeletal structure of an actor and the joints, while AMC files describe the actual movements of the joints in an actor's skeletal structure. AMC files are used in conjunction with ASF files to create a complete representation of an actor's movements, which we use for research in this paper.

For any point $(x,y,z)$ in the Cartesian coordinate system, it can be represented in the spherical coordinate system using $(r,\theta,\phi)$. The equation for converting the Cartesian coordinate system to the spherical coordinate system is given as:%引用球坐标系
\begin{align}
	\begin{cases}
	r=\sqrt{x^{2}+y^{2}+z^{2}}\\
        \theta=\arccos{(\frac{z}{r})} \\
        \phi=\arctan{(\frac{y}{x})}
	\end{cases}
        \label{trans_01}
\end{align}

We then assume that $\vec{v}_c$ and $\vec{v}_s$ is the velocity of the point in the Cartesian coordinate system and in the spherical coordinate system. According to the operation of the vector, $\vec{v}_c$ and $\vec{v}_s$ can be represented as:
\begin{equation}\label{vc}
        \vec{v}_c=\dot{x}+\dot{y}+\dot{z} 
\end{equation}
\begin{equation}
        \vec{v}_s=\dot{r}+\dot{\theta}+\dot{\phi}  
\end{equation}

Since in the Cartesian coordinate system, the vector can be decomposed according to different orthogonal unit vectors. $\vec{v}_c$ can also be decomposed according to the direction of $(r,\theta,\phi)$, which is represented as $(\widehat{r},\widehat{\theta},\widehat{\phi})$. Taking the relationship between angular velocity and linear velocity into consideration, we derive the equations of the velocity:
\begin{equation}      
        \vec{v_c}=\dot{r}\widehat{r}+r\dot{\theta}\widehat{\theta}+r\sin{\theta}\dot{\phi}\widehat{\phi}
        \label{trans_02}
\end{equation}

Combined (\ref{trans_01}) (\ref{vc})with (\ref{trans_02}), we can obtain the first derivative of r, $\theta$ and $\phi$ as:
\begin{equation}      
\dot{r}=\sin{\theta}\cos{\phi}\dot{x}+\sin{\theta}\sin{\phi}\dot{y}+\cos{\theta}\dot{z}\quad
\label{dot_r}
\end{equation}
\begin{equation}      
\dot{\theta}=\frac{\cos{\theta}\cos{\phi}}{r}\dot{x}+\frac{\cos{\theta}\sin{\phi}}{r}\dot{y}-\frac{\sin{\theta}}{r}\dot{z}
\label{dot_theta}
\end{equation}
\begin{equation}      
\dot{\phi}=-\frac{\sin{\phi}}{r\sin{\theta}}\dot{x}+\frac{\cos{\phi}}{r\sin{\theta}}\dot{y}\qquad\qquad\qquad\quad
\label{dot_phi}
\end{equation}

As the bone length is fixed for a human, r is a constant that represents the bone length. The spherical coordinate system with a fixed r can also be called spherical polar coordinates. From $\dot{r}$=0, The equations \ref{dot_theta} and \ref{dot_phi} can be simplified as:
\begin{align}
    \begin{cases}
        \dot{\theta}=-\frac{\dot{z}}{r\sin{\theta}}\\
        \dot{\phi}=\frac{\dot{y}\sin{\theta}+\dot{z}\cos{\theta}\sin{\phi}}{r\sin^{2}{\theta}\cos{\phi}}
    \end{cases}    
    \label{trans_final}
\end{align}

By using the coordinate system transformation equations \ref{trans_final}, we can convert all the points in a sequence from the Cartesian coordinate system to the spherical coordinate system. We denote the sequence as $ S=\{F_n\big|  1\leq n \leq N  \} $, where $F_n$ is the $n^{th}$ frame in the sequence, and $N$ is the total number of the frames. Subsequently, for each frame, $F_n$, each point of the frame can be represented by $\theta$ and $\phi$. We can  further denote $ F_n=\{(\theta_{n,m},\phi_{n,m}) \big| 1\leq m \leq M \}$, where $(\theta_{n,m},\phi_{n,m})$ is the $m^{th}$ point in the frame and $M$ is the the amount of points in each frame. Moreover, the position of the points is also related to the length of the bones, which can be denoted as $R = \{ r_m \big| 1\leq m \leq M  \}$. Since $R$ represents the length of the bones, it is fixed in a given sequence. So, $R$ is not included during reconstruction.

For two frames $F_{0}$ and $F_{K}$, we assume that the overall movement follows polynomial functions:
\begin{align}
\left\{
	\begin{aligned}
	\theta_{t}=\sum_{\lambda=0}^{\lambda^{max}}A_{\lambda}t^{\lambda}\\
	\phi_{t}=\sum_{\lambda=0}^{\lambda^{max}}B_{\lambda}t^{\lambda}
	\end{aligned}
\right.
\end{align}

We set the frame $F_{0}$ as the frame in time $t_0$ and $F_{K}$ as the frame in time $t_K$. For the $m^{th}$ point we now have the information $(\theta_{0,m},\phi_{0,m})$ and $(\theta_{K,m},\phi_{K,m})$. Also, we get the velocity information $(\dot{\theta}_{0,m},\dot{\phi}_{0,m})$ and $(\dot{\theta}_{K,m},\dot{\phi}_{K,m})$. Then, four equations can be obtained by inputting the values into the function for each parameter. So we set the highest power $\lambda^{max} = 3 $, and the function can be solved using the location and velocity information of $F_{0}$ and $F_{K}$, which can be denoted as:
\begin{align}
\left\{
	\begin{aligned}
	\Theta_m(t)=\sum_{\lambda=0}^{3}A_{m,\lambda}t^{\lambda}\\
	\Phi_m(t)=\sum_{\lambda=0}^{3}B_{m,\lambda}t^{\lambda}
	\end{aligned}
\right.
\label{rec_pe}
\end{align}

As the time interval between two adjacent frames should be the same in a sequence, we denote the time interval as $\Delta$. The $m^{th}$ of the $k^{th}$ frame in the reconstructed sequence can be derived by:
\begin{align}
\left\{
	\begin{aligned}
	\Tilde{\theta}_{k,m}=\Theta_m(t_0+k\Delta)\\
	\Tilde{\phi}_{k,m}=\Phi_m(t_0+k\Delta)
	\end{aligned}
\right.
\end{align}

After reconstruct all the sections in the sequence, we can get the new sequence $ \Tilde{S}=\{\Tilde{F}_n \big|  1\leq n \leq N  \} $, where $ \Tilde{F}_n=\{(\Tilde{\theta}_{n,m},\Tilde{\phi}_{n,m}) \big| 1\leq m \leq M \}$.

To determine the location of each point using the spherical coordinate system, we need to know the coordinates of the root point in the Cartesian coordinate system. Since the root point does not have an origin point, it can not be determined using the spherical coordinate system. This also means that the root point will not involve changes in the bone length during reconstruction, which makes the interpolation algorithm in the Cartesian coordinate system a good choice. Also, compared with all the other points, the root point is the most stable point. As a result, the trajectory of the root point can be well reconstructed with polynomial interpolation in the Cartesian coordinate system.

Similarly, we first denote the coordinate of the root point in frame $F_{0}$ and $F_{K}$ as $O_0$ and $O_K$. We also have the velocity of the root point, which is shown as $\dot{O}_0$ and $\dot{O}_K$. By substituting the four values, we can calculate the parameter of another polynomial equation:
\begin{equation}
    \Gamma(t)=\sum_{\lambda=0}^{3}C_{\lambda}t^{\lambda}
\end{equation}
from which the position of the root point of the $k^{th}$ frame can be obtained via:
\begin{equation}
    	O_k=\Gamma(t_0+k\Delta)
\end{equation}

%Note to Paul himself - read the above tmr
\subsection{Keyframe Extraction}

Keyframe extraction of motion capture is one of the techniques used to determine the most critical frames in a motion sequence. The extracted keyframes can be used to achieve various goals, e.g., summarize the sequence for the users using several frames and enable efficient storage. More importantly, by combining keyframe extraction and reconstruction methods, we can significantly decrease the bandwidth requirement, which can be used to handle the network congestion and decrease the latency. 

Most existing work chooses the labeled keyframes of the ground truth by humans. However, such a kind of ground truth keyframes is not the optimal decision for reconstruction. To achieve near-optimal reconstruction performance, we want to find an intelligent way to extract the proper keyframes for the proposed reconstruction algorithm. The keyframes extraction problem considering the reconstruction process will be elaborated in this section.

For each given sequence $ S=\{F_n\big|  1\leq n \leq N  \} $, we denote the keyframe extraction decision as $ K=\{k_w\big|  1\leq w \leq W  \} $, where $k_w$ is the index of the $w^{th}$ keyframe and $W$ is the total number of extracted keyframes. For the section between the $w^{th}$ keyframe and the $(w+1)^{th}$ keyframe, we apply the reconstruction algorithm and calculate out the polynomial equation $\Theta_m^w(t)$ and $\Phi_m^w(t)$ for the $m^{th}$ point. 

For all the points, we can use the same process and derive the trajectory equations $\Theta^w(t)=\{\Theta_1^w(t),\Theta_2^w(t),\cdots,\Theta_M^w(t)\}$ and $\Phi^w(t)=\{\Phi_1^w(t),\Phi_2^w(t),\cdots,\Phi_M^w(t)\}$ for the $w^{th}$ section. Then for any point in the $q^{th}$ reconstructed frame, the position can be obtained by:
\begin{align}
\left\{
	\begin{aligned}
	\Tilde{\theta}_{k_w+q,m}=\Theta_m^w(t_{k_w}+q\Delta)\\
	\Tilde{\phi}_{k_w+q,m}=\Phi_m^w(t_{k_w}+q\Delta)
	\end{aligned}
\right.
\end{align}
where $t_{k_w}$ represents the timestamp of the $w^{th}$ keyframe. Therefore the total error between the original frames and the new frames in this section for $\theta$ and $\phi$ can be calculated by:
\begin{equation}
\left\{
        \begin{aligned}
        E_\theta^w=\sum_{q=0}^{\Delta k} \sum_{m=1}^M\big|\Theta_m^w(t_{k_w}+q\Delta)-\theta_{k_w+q,m}\big|\\
        E_\phi^w=\sum_{q=0}^{\Delta k} \sum_{m=1}^M\big|\Phi_m^w(t_{k_w}+q\Delta)-\phi_{k_w+q,m}\big|
        \end{aligned}
        \right.
\end{equation}
where $E_\theta^w$ is the total error of $\theta$, $E_\phi^w$ is the total error of $\phi$ and $\Delta k$ is the frames number in the section. So, the total error of section $w$ can be given as:
\begin{equation}
    E^w=E_\theta^w+E_\phi^w
\end{equation}

We repeat the same reconstruction process for all the sections decided by the keyframe extraction decision $K$. The overall trajectory equations are derived, which can be written as $\Theta(t)=\{\Theta^1(t),\Theta^2(t),\cdots,\Theta^{W-1}(t)\}$ and $\Phi(t)=\{\Phi^1(t),\Phi^2(t),\cdots,\Phi^{W-1}(t)\}$. After that, we combined each section into the constructed new sequence $ \Tilde{S}=\{\Tilde{F}_n \big|  1\leq n \leq N  \} $, where $ \Tilde{F}_n=\{(\Tilde{\theta}_{n,m},\Tilde{\phi}_{n,m}) \big| 1\leq m \leq M \}$. 

Thereafter, the mean error between $ S $ and $\Tilde{S} $ is denoted as $E$. Since the keyframes will be the same during reconstruction, the total error can be derived by summing the error of each section. Then we have the function:
\begin{equation}
    E=\frac{1}{N}\frac{1}{M}\sum_{w=1}^{W-1}E^w=\frac{1}{N}\frac{1}{M}\sum_{w=1}^{W-1}(E_\theta^w+E_\phi^w)
\end{equation}

Since that, the points are generated by $\Theta(t)$ and $\Phi^w(t)$, which can be determined based on sequence $S$ and the keyframe extraction decision $K$. Other parameters, such as point number $M$, frames number $N$, keyframe number $W$, and the time interval between frames $\Delta$ are fixed hyperparamter, which are given by the application environment. Resulting from that, the mean error can be determined by $S$ and $K$. We further defined the error function $Q$, which can be formulated as:
\begin{align}
    Q(S,K)=&\frac{1}{N}\frac{1}{M}\sum_{w=1}^{W-1}\sum_{q=0}^{\Delta k} \sum_{m=1}^M(\big|\Theta_m^w(t_{k_w}+q\Delta)-\nonumber\\
    &\theta_{k_w+q,m}\big|+\big|\Phi_m^w(t_{k_w}+q\Delta)-\phi_{k_w+q,m}\big|)
    \label{Q}
\end{align}

From the error function $Q(S,K)$, we can see that for the given sequence $S$, the reconstruction error can be calculated by the keyframe extraction decision $K$. Therefore, our target for finding the best keyframes for the proposed reconstruction algorithm can be transformed into an optimization problem ($\mathcal{P}$):
\begin{subequations}
\begin{equation}
    \quad (\mathcal{P}) \quad min: Q(S,K) \qquad \qquad \qquad \qquad
\end{equation}
\begin{equation}
    s.t.: k_w\in \mathbb {Z}, 1\leq w \leq W \quad
\end{equation}
\begin{equation}
    \quad\qquad1<k_w<N, 1\leq w \leq W
\end{equation}
\begin{equation}
   \qquad\qquad k_{i}\neq k_{j}, 1\leq i,j \leq W, i\neq j
\end{equation}
\end{subequations}

The problem ($\mathcal{P}$) is an optimization problem with a very large search space and is hard to solve by traditional optimization algorithms. To solve this problem with efficiency, we proposed a new machine learning algorithm using DQN. The next section elaborates on the details. %which will be elaborated on in the next section.

\section{SIDQL ALGORITHM}

\begin{figure*}[htbp]
	\centerline{
	\includegraphics[scale=0.065,trim=0in 0.1in 0in 0.4in]{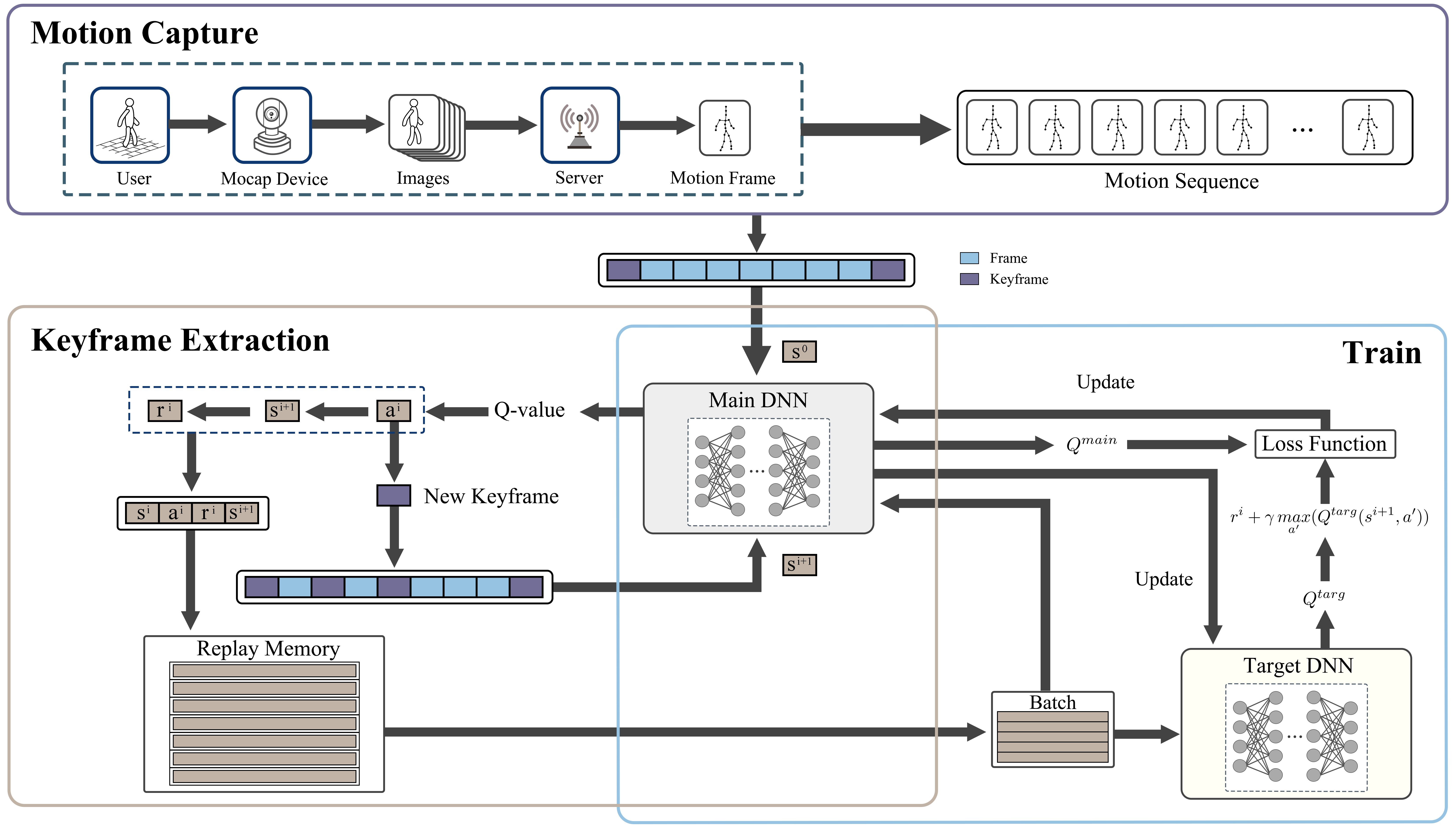}}
	\caption{The procedure of SIDQL.}
	\label{Frame_algorithm}
\end{figure*}

This section elaborates on our algorithm for solving the problem ($\mathcal{P}$). The sequences are transformed to the spherical coordinate system, and a DQN is used to extract the keyframes. A special reward function is also proposed to further ensure the accuracy of the proposed reconstruction approach. Fig \ref{Frame_algorithm} shows the framework of the whole process.

\subsection{Framework}
For any sequence $ S $, we first convert the location data $(x,y,z)$ and velocity data $\vec{v}_c$ into the spherical coordinate system using \ref{trans_01} and \ref{trans_final}. Since that $r$ is constant for the sequence, the input data of the $n^{th}$ frame can be represented by $(F_n,V_n)$, where $ F_n=\{(\theta_{n,m},\phi_{n,m}) \big| 1\leq m \leq M \}$ and $ V_n=\{(\dot{\theta}_{n,m},\dot{\phi}_{n,m}) \big| 1\leq m \leq M \}$. Considering the input format of DQN, we denote the keyframe extraction decision $K$ as a $\mathcal{X}=\{x_1,x_2,\dots,x_N\}$, which can be defined by:
\begin{equation}
x_n= 
\begin{cases}
1, & \textrm{if the $n^{th}$ frame is keyframe}, \\
0, & \textrm{if the $n^{th}$ frame is not keyframe}. 
\label{decision_change}
\end{cases}
\end{equation}

Deep Q-Learning is one of the typical algorithms of DRL, which can be used to solve the decision-making problem of a Markov Decision Process (MDP). To apply the DQL algorithm, we regard the keyframe extraction process as a MDP process. For each time step, we add one new keyframe to $\mathcal{X}$ until we reach the desired keyframe number.

We represent $s^i=\{F^i,V^i,\mathcal{X}^i\}$ is the state when extracting the $i^{th}$ new keyframe. To meet the prerequisite of reconstruction, we set the first frame and the last frame as two initial keyframes. For the sequence with $N$ frames, the actions space contains $N$ different actions. Specifically, the action $a^i=m$ means at state $s^i$, we add the $m^{th}$ frame to $\mathcal{X}$ as the new keyframe. Additionally, the keyframe is chosen from the non-keyframes. 

Although all keyframes should be treated fairly and the keyframes should be extracted simultaneously, DQN has the ability to generate actions for maximizing long-term reward. As a result, the impact of converting the decision-making process into iterations can be ignored. Theoretically, the reconstructed sequence has the lowest accuracy at state $s^0$, where all the middle frames are reconstructed from the first frame and the last frame. When all the frames are considered keyframes, the reconstructed sequence is the sequence $S$ itself. We denote the reconstruction in the $i^{th}$ state as $Q^i$, which can be calculated by (\ref{Q}). We then set a normalized relative error $R^i$ of $s^i$ as:
\begin{equation}
    R^i=1-\frac{Q^i}{Q^0}
\end{equation}
where $R^i$ is more close to one if $s^i$ achieve higher accuracy. To determine the reward $r^i$ from each action, we compare the normalized relative error between and after the action $a^i$. Then $r^i$ can be denoted as:
\begin{equation}
    r^i=R^{i+1}-R^i
    \label{reward}
\end{equation}

Consequently, the keyframe extraction process can be regarded as one MDP process represented with state $s^i$, action $a^i$ and reward $r^i$. Due to the infinite state space, considering the different positions and velocities of each point, traditional RL approaches are unable to handle this problem. DRL is then proposed by combining RL with DNN to tackle the infinite state space. Substituting different state data into DQN can output the Q-value of each possible action. Q-value is the value used to represent the scores for adapting each action. The higher the Q value, the higher the expected long-term reward after adopting the corresponding action. After the DQN is well-trained, we can determine the best keyframe extraction decision by choosing the action with the highest Q-value. 

\subsection{Train}

To train the DQN, we first collect a training dataset consisting of various sequences. When the iteration begins, we input one sequence into the model from state $s^0$. Then, in each time step for state $s^i$, we can obtain action $a^i$, reward $r^i$ and the next state $s^{i+1}$ via DQN. Then $(s^i,a^i,r^i,s^{i+1})$ can be regarded as one training data, which is stored in the training dataset. If the data number of the training dataset has reached the upper limit, the oldest data will be replaced to improve the training performance. In addition, to better explore the possible keyframe extraction decisions, we also apply $\epsilon$-greedy policy in the decision-making process. Within each step, a random number is generated as a determination. Specifically, $a^i$ will be chosen randomly if the determination is smaller than $\epsilon$. Otherwise, $a^i$ will be generated using DQN as planned.

After the training dataset has got sufficient training data, the DQN will be trained every $\tau$ step, namely the training interval. To train the DQN during each training phase, we randomly select one batch of data from the dataset. Since the main goal of DQL is to maximize the reward of the final decision rather than focusing on the reward for each step, two DNNs with the same structure are used in one DQL model. One is the main DNN used to output the Q value of the current state, while another is the target DNN utilized to evaluate the next state decided by the main DNN. The Q-value produced by the main DNN is called $Q^{main}$, whereas the Q-value of the target DNN is denoted as $Q^{targ}$. For training data $(s^i,a^i,r^i,s^{i+1})$, the expected output of $Q^{main}(s^i,a^i)$ can be given as:
\begin{equation}
    r^i+\gamma \mathop{max}\limits_{a'} (Q^{targ}(s^{i+1},a’))
\end{equation}
where $\gamma$ is the discount factor that balances the short view and the future reward. For all the sequences in the batch, we apply the Huber function as the loss function to calculate the loss between the output Q-value and the expected Q-value, which is then used to update the parameter of the main network.

After the model is trained, the performance of $Q^{main}$ becomes better for extracting a new keyframe. The ability of $Q^{targ}$ then limits the training efficiency and accuracy of the model. Consequently, after $\pi$ epochs of training, we renew the network parameter of $Q^{targ}$ using the trained parameter of $Q^{main}$. The whole model can be trained with stability and efficiency as the iteration continues. The process is repeated until the model converges. The pseudo-code of the proposed SIDQL algorithm is provided in Algorithm \ref{algorithm_code}.

\begin{algorithm}[htbp]
\caption{SIDQL Algorithm} %

\begin{algorithmic}[1]
%{\bf Input:} Training dataset $\mathcal{S}$, sequence $S^*$ and keyframe number W\\
%{\bf Output:} Keyframe extraction decision K
\Require Training dataset $\mathcal{S}$, sequence $S^*$ and keyframe number W
\Ensure Keyframe extraction decision K

\State \textbf{Initialization:} Initialize the main network $Q^{main}$, the target network $Q^{targ}$ and the replay memory
\State Training steps j = 0
\For { epochs i $\leq$ I}
    \State Choose a sequence $S^i$ from $\mathcal{S}$
    \State Initialize the keyframe extraction decision $K_0$
    \For{w $\leq$ W}
        \State Regard $S^i$ and current decision $K_w$ as state $s^i$
        \State j = j + 1
        \State Generate a random number p within [0,1]
            \If {p $\leq$  $\epsilon$ }
                \State Randomly choose action $a^i$
            \Else
                \State Set $a^i$=arg$\mathop{max}\limits_{a^i}$ $Q^{main}(s^i,a^i)$
            \EndIf
            \State Calculate the reward $r^i$ using (\ref{reward})
            \State Renew decision $K_w$ according to $a^i$
            \If {database is not full}
                \State Store $(s^i,a^i,r^i,s^{i+1})$ in the replay memory
            \Else
                \State Replace the oldest data with $(s^i,a^i,r^i,s^{i+1})$
            \EndIf
            \If{ mod(j,$\tau$)==0}
                \State Randomly choose one batch of training data
                \State Train $Q^{main}$ using the Adam optimizer
            \EndIf
            \If{ mod(j, $\tau \pi$)==0}
                \State Renew the parameter of $Q^{targ}$
            \EndIf
    \EndFor
\EndFor
\State Regard $S^*$ and decision $K_0$ as state $s^0$
\For{w $\leq$ W}
    \State Set $a^i$=arg$\mathop{max}\limits_{a^i}$ $Q^{main}(s^i,a^i)$
    \State Renew decision $K_w$ according to $a^i$
\EndFor
\State \Return 
Keyframe decisions K of $S^*$ 
\end{algorithmic}
\label{algorithm_code}
\end{algorithm}

\subsection{Test}

To verify the performance of the proposed algorithm, we first need to ensure that the model has converged after iterations. The loss values during each training epoch can be utilized to test the convergence performance. More importantly, we also need to evaluate the ability to solve the problem after the DQN model is converged.

The error between the generated and optimal decisions is hard to measure because it is hard to obtain the optimal solutions for the proposed keyframe extraction problem ($\mathcal{P}$). So, we choose several algorithms as baselines and compare analyses on the same test dataset. We assume that $T$ different sequences are contained in the test dataset $\mathcal{T}$, which can be denoted as $\mathcal{T}=\{S^*_t \big |1\leq t \leq T\}$. Then we define the mean angle error $\mathcal{R}$ as the evaluation criterion, given as:
\begin{equation}
    \mathcal{R}=\frac{1}{T}\sum^T_{t=1}Q(S^*_t,K^*_t)
    \label{mae}
\end{equation}
where $K^*_t$ is the keyframe extraction decision of sequence $S^*_t$ generated by different approaches.

With the mean angle error function (\ref{mae}), we can first evaluate the influence of different hyperparameter settings. The best settings can be obtained by choosing the network with the lowest error. Also, the performance under different application scenarios should be considered. The mean angle error of the proposed approach and the baselines when extracting different numbers of keyframes will be further calculated to give us more insight into setting the keyframe numbers accordingly.

\section{EXPERIMENTS}

In this section, we set up simulations to illustrate that our scheme can effectively decrease the amount of data and transmission latency, while still maintaining a high level of accuracy in reconstruction. Experiments are designed by using the motion capture data in mixed categories.

\subsection{Dataset}

We conducted several experiments on the CMU Graphics Lab Motion Capture Database~\footnote{http://mocap.cs.cmu.edu}, which is widely used in the related work of motion capture. It encompasses a diverse range of human motion data, amounting to approximately 10 hours of motion in total. In detail, the database contains 2605 sequences of motions in 6 categories and 23 subcategories by 111 human subjects, encompassing various activities such as locomotion, physical activities and interaction. By utilizing the database, we can access a large and diverse set of motions to validate the performance of the SIDQL algorithm.

\subsection{Experiment Settings}

In our experiments, we focus on utilizing 23 major joints of the human skeleton by excluding the unused joints from fingers, thumbs and toes to extract the keyframes. For each motion sequence, the frame rate reduction is applied from the initial 120 frames per second (FPS) to 30 FPS, so the time interval between two adjacent frames is 0.33s. Then, we crop the motion sequences into fixed-length motion subsequences of 60 frames each, which served as input for training the network. Subsequently, the database is divided into two subsets, where 80$\%$ of the sequences is the training set, and the rest 20$\%$ is the test set.

During the training of DQN, the agent interacts with the environment by selecting actions based on the current Q-values and receives a reward for each action taken. The parameters in the main network are updated every 200 steps, while the weights of the Q-network are updated every 100 epochs. In addition, the discount factor of the reward function is set as $\gamma$ = 0.5.

\begin{table}[htbp]   
\renewcommand{\tablename}{TABLE}
\caption{Evaluation Parameter}   
\begin{center}
\resizebox{0.6\textwidth}{!}{
\begin{tabular}{ll}    
\toprule    \bf{Parameters} & \bf{Values}\\    
\midrule  

The number of frames in a sequence & N = 60.\\
The number of points in a frame & M = 23.\\
The frames per second of the sequence & FPS = 30.\\
The time interval between two adjacent frames & $\Delta$ = 0.33s.\\
The training interval of DQN & $\tau$ = 200 steps.\\
The renew interval of $Q^{targ}$ & $\pi$ = 100 epochs.\\
The discount factor of the reward function & $\gamma$ = 0.5.\\

\bottomrule   
\end{tabular}  }
\end{center}
\label{settings}
\end{table}

For the DQN in our algorithm, we consider the main network and Q-network with two hidden layers as the decision-making module and a fully connected network to extract the features of the keyframes. We implement our algorithm in Python 3.9.12 with TensorFlow 3.6.13. All the simulations are performed based on an Intel Core i9-12900H CPU and 32.0 GB memory.

\subsection{Training Evaluation}

In this part, we evaluate the performance of the SIDQL algorithm by measuring the mean error and loss value under different learning rates, training intervals, memory sizes, and batch sizes.

\subsubsection{Impact of Learning Rate}
\ 
\newline
\indent Learning rate plays a vital role as a hyperparameter that significantly impacts the performance of our approach. In DQN-based models, the learning rate has a crucial influence. When the learning rate is too low, the model converges slowly, requiring more epochs for training or getting stuck in suboptimal solutions. Conversely, if the learning rate is too large, the model converges quickly but may fail to achieve high accuracy due to overshooting or instability.

\begin{figure}[htbp]
	\centerline{
	\includegraphics[scale=0.13]{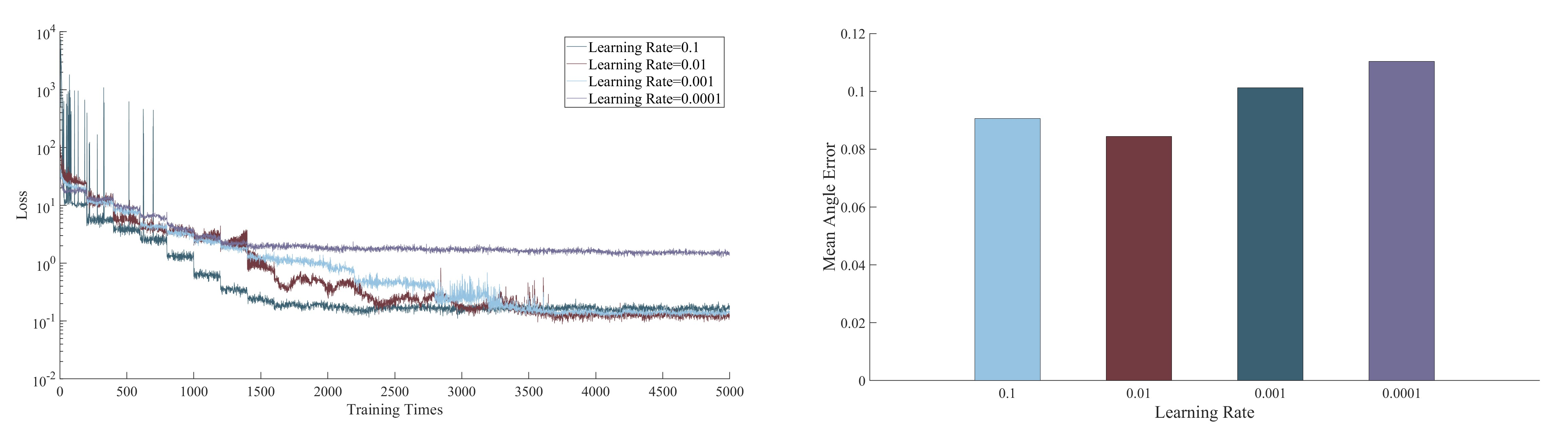}}
	\caption{Simulation result of learning rate. Left: Loss performance of learning rate. Right: Mean angle error of learning rate.}
	\label{LR}
\end{figure}

During our experiments, the learning rate is adjusted from 0.1 to 0.0001, and the loss performance can be shown in the left image of Fig. \ref{LR}. Since our algorithm is initialized with random weights, the predictions may be far from the target values at the beginning of training, resulting in a high loss. So, when plotting the figure, the logarithm base 10 is applied to the loss values to better visualize and interpret the data. It is observed that the loss generally converges to a stable value regardless of the learning rate used. The changing speed of the loss values becomes faster at the beginning and then becomes stable when the algorithm starts to learn and adjust its parameters. When our algorithm updates the weights of the network, the loss decreases significantly. Moreover, a learning rate of 0.01 can lead to a stable and effective training process with minimized loss. An overly high learning rate may result in fluctuating loss values, indicating an unstable training process. Conversely, a learning rate that is too low may lead to slow convergence or getting stuck in suboptimal solutions, which hinders the model's capacity to extract accurate and representative keyframes.

Indeed, the loss values can affect the quality of the reconstructed motion. Fig. \ref{LR} (right-hand side) shows the mean angle error under different learning rates. When the learning rate is set to 0.01, it yields the minimum loss value and the lowest error, suggesting that the algorithm achieves the best reconstruction performance in minimizing the discrepancy between the predicted motions and the ground truth.

\subsubsection{Impact of Training Interval}
\ 
\newline
\indent Training interval in our algorithm refers to the frequency at which the main network updates its parameters through training. It determines how often the model learns from its experiences and adjusts its action-selection strategy. By setting an appropriate training interval, we can balance efficiency and stability in our algorithm.

\begin{figure}[htbp]
	\centerline{
	\includegraphics[scale=0.105]{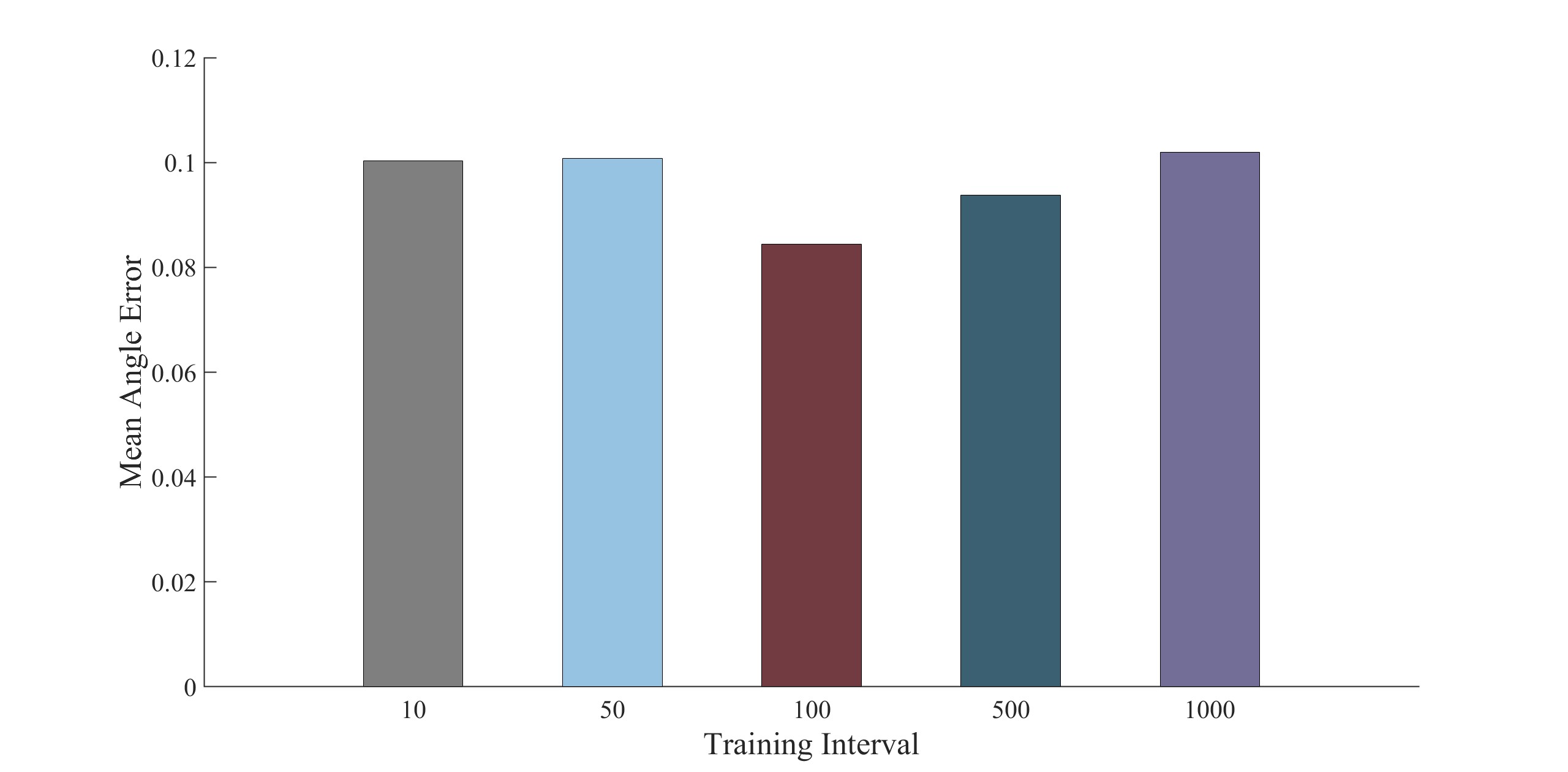}}
	\caption{Mean angle error of training interval.}
	\label{TI_error}
\end{figure}

Training interval determines how often the model parameters are updated. Still, it does not directly influence the calculation of the loss value, so we do not analyze the loss performance under different training intervals. Fig. \ref{TI_error} shows the mean angle error of different training intervals in motion reconstruction. It is important to note that training intervals cannot be too large or too small. On the one hand, if the training interval is too small, meaning that the model is updated frequently, it can lead to overfitting. On the other hand, if the training interval is too large, the model updates infrequently, which can slow down the learning process, resulting in a large computational cost in the keyframe extraction process. Consequently, we set the training interval to 100 to avoid overfitting and more training time to extract the keyframes.

\subsubsection{Impact of Memory Size}
\ 
\newline
\indent Memory size, often referred to as the replay buffer size, is an important factor in DQN-based models due to its impact on the learning process and the model's ability to effectively utilize past experiences. For traditional DQN-based models, if the memory size is too large, the model may spend more time replaying redundant or less informative experiences, leading to slower convergence and inefficient learning. On the contrary, the model is more likely to sample experiences that are temporally correlated or biased towards recent data, compromising the model's performance. 

\begin{figure}[htbp]
	\centerline{
	\includegraphics[scale=0.19]{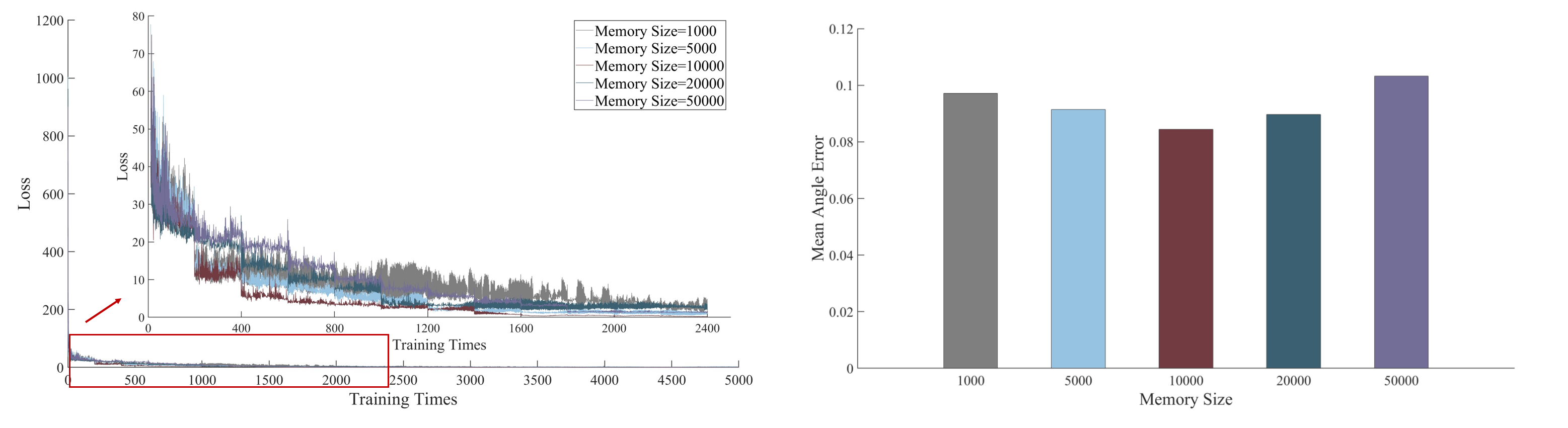}}
	\caption{Simulation result of memory size. Left: Loss performance of memory size. Right: Mean angle error of memory size.}
	\label{MS}
\end{figure}

During our experiments, memory size is adjusted from 1000 to 50000, and each model is trained with 5000 training steps, as shown in the left image in Fig. \ref{MS}. Obviously, irrespective of the memory size used, the loss typically converges to a stable value during the training process. When memory size is relatively small, the loss exhibits significant fluctuations, potentially affecting the accuracy and reliability of the extracted keyframes. On the contrary, with a large memory size, the model introduces computational inefficiencies and redundant data. So, striking the right balance in memory size is crucial for achieving accurate and efficient keyframe extraction in motion capture. When the memory size is set to 10,000, the loss reaches its minimum and remains stable. Additionally, we can notice that the loss tends to stabilize after around 1600 training iterations. This stabilization indicates that the model has reached a relatively stable state in terms of its learning progress, which can positively impact the accuracy and reliability of keyframe extraction.

The right image of Fig. \ref{MS} presents the mean angle error across various memory sizes, providing insights into the impact of memory size on the accuracy of the motion reconstruction process. It is evident that the mean angle error is minimized when the memory size is set to 10,000, indicating that the keyframe extraction decisions are optimal at this memory size. Therefore, selecting an appropriate memory size is crucial for achieving superior results in motion capture applications.

\subsubsection{Impact of Training Batch Size}
\ 
\newline
\indent Batch size is another hyperparameter that can be tuned to improve the performance of DQN-based models, which is the number of experiences (i.e., state, action, reward, next state) that are fed into the network at once during training. If the batch size is too small, the gradient estimates used to update the weights of the network will be noisy and may not accurately reflect the true gradient of the loss function, resulting in slow convergence, poor performance and instability in the learning process. On the other hand, if the batch size is too large, the network may become overfitted. It will influence the generalization ability of the model.

\begin{figure}[htbp]
	\centerline{
	\includegraphics[scale=0.19]{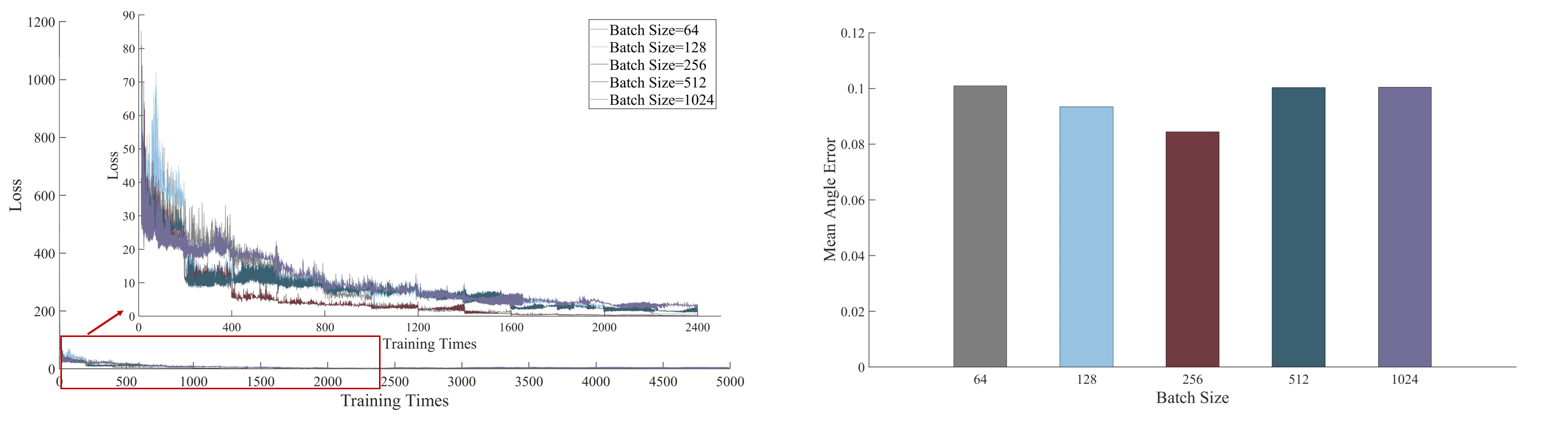}}
	\caption{Simulation result of batch size. Left: Loss performance of batch size. Right: Mean angle error of batch size.}
	\label{BS}
\end{figure}

The left image of Fig. \ref{BS} illustrates the loss performance obtained during our experiments, where we vary the batch size from 64 to 1024. Choosing an appropriate batch size that balances the trade-off between stability and efficiency in our algorithm is important. The minimum loss is achieved and maintained by setting the batch size to 256. It is worth noticing that when the batch size is small, the loss shows significant fluctuations, and the fluctuation is also significantly larger than the big batch size after the loss convergence. That is because when the batch size is too small, our algorithm may not be able to effectively learn the features of the input motion data, so the extracted keyframes cannot accurately represent the sequence. But when the batch size is too large, it can lead to overfitting, also resulting in poor performance in keyframe extraction. In addition, no matter what batch size is used, the loss usually converges to a stable value during the training process.

The mean angle error across various batch sizes is shown in the right image in Fig. \ref{BS}. Obviously, the mean angle error is minimized when the batch size is set to 256, showing that our algorithm can successfully learn the input motion data's features at this batch size and prevent overfitting. Additionally, the mean angle error is nearly identical when the batch size is 512 and 1024. This can be attributed to the fact that a larger batch size can cause slower convergence or overfitting, which can adversely affect the performance of motion reconstruction after selecting the keyframes.

\subsection{Performance Comparison}

By evaluating the mean error and loss performance under different hyperparameters, we have proven that our SIDQL algorithm can be trained without labeled data and effectively extract the keyframes. To better understand our algorithm's performance, we choose several traditional algorithms as baselines. We then changed the number of keyframes from 5 to 15 and tested the mean error of the methods. The following methods are tested for comparison analysis:

\begin{itemize}
\item \emph{Random Choice (RC)}: In this method, we first select the first and last frames of the motion sequence as keyframes, then randomly select a certain number of keyframes from the other frames. Each frame has an equal probability of being selected as a keyframe.
\item \emph{Uniform Choice (UC)}: In this method, the first and last frames of the motion sequence are selected as keyframes. We then uniformly select frames from the sequence as keyframes.
\item \emph{Greedy}: This method involves setting an original keyframe set that includes the first and last frames of a given motion sequence. For each frame in a motion sequence, we set the current frame and other keyframes in the original keyframe set as the new keyframe set, then apply the reconstruction algorithm and calculate the mean error of the new keyframe set. Subsequently, we select the frame with the lowest mean error as the next keyframe, and add it to the original keyframe set. The iteration is repeated the desired number of keyframes is selected. Since the Greedy algorithm explores multiple possibilities for extracting keyframes, it achieves high accuracy. Hence, it is defined as the skyline.
\item \emph{Our algorithm}: In this method, we applied the proposed SIDQL algorithm to extract the keyframes.
 \end{itemize}

\begin{table}[htbp]   
\renewcommand{\tablename}{TABLE}
\caption{Mean Error under different number of keyframes}   
\label{Mean Error}
\begin{center}
\resizebox{0.6\textwidth}{!}{%
\begin{tabular}{cccc}    
\toprule    \bf{Method} & \bf{5 keyframes} & \bf{10 keyframes} & \bf{15 keyframes}\\ 
\midrule  RC & 0.1437 & 0.0833 & 0.0549\\ 
 UC & 0.0945 & 0.0490 & 0.0328 \\  
 Greedy & 0.0536 & 0.0311 & 0.0200 \\
 \textbf{Ours} & 0.0844 & 0.0443 & 0.0297\\
\bottomrule   
\end{tabular}  }
\end{center}
\end{table}

%\begin{figure*}[htbp]
	%\centering
	%\subfloat[]{\includegraphics[width=1.8in]{figure/result/jump.png}}%
		%\label{(a)}
	%\hfil
	%\subfloat[]{\includegraphics[width=1.8in]{figure/result/run.png}}%
		%\label{(b)}
        %\hfil
	%\subfloat[]{\includegraphics[width=1.825in]{figure/result/pick.png}}%
		%\label{(c)}
	%\caption{The reconstructed motion by different methods when extracted 5 keyframes. (a) Jump. (b) Run. (c) Pick up something.}
	%\label{motion1}
%\end{figure*}

\begin{figure}[htbp]
	\centerline{
	\includegraphics[scale=0.265]{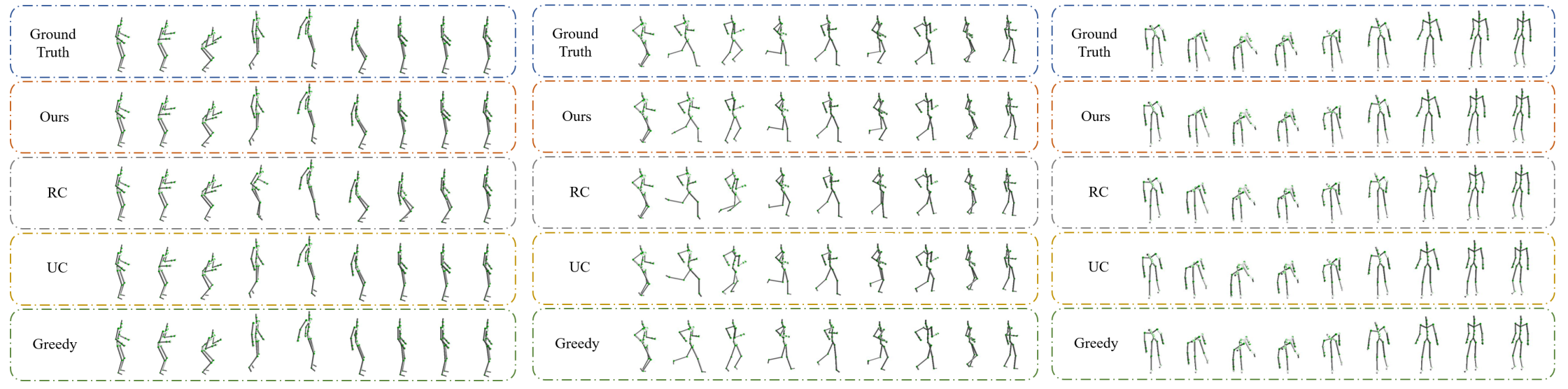}}
	\caption{The reconstructed motion by different methods when extracted 5 keyframes. Left: Jump. Middle: Run. Right: Pick up something.}
	\label{motion}
\end{figure}

The comparison results of different baselines are depicted in Table \ref{Mean Error}. In comparison between the RC and UC methods, our method extracts better keyframes with less reconstruction error in different numbers of keyframes. In the RC method, randomly selecting keyframes may lead to the omission of crucial keyframes, which can negatively impact the accuracy of motion reconstruction. For the UC method, when the motion changes rapidly, the reconstructed motion sequence is not coherent, which affects the user experience. Our method selects keyframes related to the motion content, improving the accuracy of keyframe extraction and reducing errors. However, our results are slightly inferior to the Greedy method. It can be attributed to the fact that the Greedy algorithm explores multiple possibilities for extracting keyframes, while our method does not consider all possible combinations. Although our method's accuracy is not as good as the Greedy method, when extracting 5 keyframes for a motion sequence, the Greedy method takes about 0.9s to generate keyframe decisions, while we can generate decisions within 0.004s. Therefore, compared to the Greedy method, our method generates decisions in a very short time, which can meet the low-latency requirements for avatar motion synchronization in the Metaverse.

We visualize the reconstruction results obtained by extracting 5 keyframes. From these animated characters, it is obvious that the reconstructed frames by our SIDQL algorithm and Greedy method are very similar to the ground truth frames, while the reconstructed frames by the RC and the UC are more or less unnatural and incoherent. In the left image of Fig. \ref{motion}, the motion sequence about jumping is from subject 01\_02. Obviously, some key joints, such as the knee, leg and elbow, move widely, and the reconstructed motion by the RC method has a significant deviation, compared to the original motion. The motion sequences reconstructed by UC, Greedy, and our method are almost indistinguishable from the ground truth sequence. When the motion is fast and wide, the differences between the motion sequences reconstructed by different methods become more apparent. For instance, the middle image in Fig. \ref{motion} displays the running motion on subject 104\_122, which is a relatively fierce motion. The motions reconstructed by the RC and the UC are not performed well. For the RC method, the motion of the arms and legs is highly uncoordinated, and there is a significant positional deviation compared to the original motion. Unlike the ground truth, the UC method's body swings as it runs and has unnatural leg movements. In contrast, the motion sequences reconstructed by the Greedy method and our method have higher accuracy, and are similar to the original motion, which indicates that the keyframes selected by these two methods are representative of the motion sequence. However, due to the fast and large movements during running, there are brief moments of unnatural motion in the foot movement of our method and the arm movement of the Greedy method. In addition, when the motion speed is slow, the differences between the motion sequences reconstructed by different methods are small, and the motion is natural and coherent, as shown in the right image of Fig. \ref{motion}.

\section{CONCLUSION AND FUTURE WORK}\label{conclusion}

In this study, we initially developed a novel motion reconstruction algorithm predicated on spherical interpolation techniques. This algorithm transforms the positional and velocity data of each point into spherical coordinates, ensuring the constancy of bone length. Building upon this innovative reconstruction method, we formulated the keyframe extraction problem as an optimization task aimed at minimizing the mean reconstruction error. Subsequently, we introduced the SIDQL framework, which incorporates a specialized reward function derived from the mean error metric. This framework is adept at learning from mixed-category motion sequences without the necessity for labeled keyframes. We then assessed the reconstruction efficacy of our proposed algorithm by comparing it with several baseline methods across varying number of keyframes. Experimental outcomes attest to our algorithm's capability to considerably curtail data volume while preserving high reconstruction fidelity, fulfilling the stringent low-latency demands of the Metaverse.

The principal contribution of this work is the proposition and validation of the SIDQL algorithm for keyframe extraction and motion reconstruction within the realm of motion capture. The algorithm we propose can be seamlessly integrated with edge computing and federated learning paradigms to further diminish transmission latency and enrich the user experience. To more efficaciously tackle real-world applications, it is imperative to conduct a comparative performance evaluation of deep learning and reinforcement learning methodologies within the same framework to identify the most efficacious strategy. Additionally, the application of meta-learning strategies can be explored to enhance the model's adaptability across diverse scenarios. Looking ahead, we aim to deploy the algorithm in tangible real-world settings.

\bibliographystyle{ACM-Reference-Format}
\bibliography{ref}

%\appendix

\end{document}